\documentclass[times,twocolumn,final]{elsarticle}

\usepackage{tabularx}

\usepackage{custom}
\usepackage{framed,multirow}
\usepackage{booktabs}
\usepackage{amssymb}
\usepackage{latexsym}

\usepackage{url}
\usepackage{xcolor}
\usepackage{amsmath}
\usepackage{hyperref}
\usepackage{makecell}
\definecolor{newcolor}{rgb}{.8,.349,.1}

\begin{document}

\begin{frontmatter}

\title{Deep learning models for gastric signet ring cell carcinoma classification in whole slide images}
\author[1]{Fahdi \snm{Kanavati}}
\author[2]{Shin \snm{Ichihara}}
\author[3]{Michael \snm{Rambeau}}
\author[3]{Osamu \snm{Iizuka}}
\author[4]{Koji \snm{Arihiro}}
\author[1,3]{Masayuki \snm{Tsuneki} \corref{cor1}}

\address[1]{Medmain Research, Medmain Inc., Fukuoka, 810-0042, Japan}
\address[2]{Department of Surgical Pathology, Sapporo Kosei General Hospital, Hokkaido, 060-0033, Japan}
\address[3]{Medmain Inc., Fukuoka, 810-0042, Japan}
\address[4]{Department of Anatomical Pathology, Hiroshima University Hospital, Hiroshima, 734-0037, Japan}

\cortext[cor1]{Corresponding author: tsuneki@medmain.com
Tel.: +81-92-707-1977; Fax: +81-92-707-1978;}

\begin{abstract}
Signet ring cell carcinoma (SRCC) of the stomach is a rare type of cancer with a slowly rising incidence. It tends to be more difficult to detect by pathologists mainly due to its cellular morphology and diffuse invasion manner, and it has poor prognosis when detected at an advanced stage. Computational pathology tools that can assist pathologists in detecting SRCC would be of a massive benefit. In this paper, we trained deep learning models using transfer learning,  fully-supervised learning, and weakly-supervised learning to predict SRCC in Whole Slide Images (WSIs) using a training set of 1,765 WSIs. We evaluated the models on four different test sets of about 500 images each. The best model achieved a Receiver Operator Curve (ROC) area under the curve (AUC) of at least 0.99 on all four test sets, setting a top baseline performance for SRCC WSI classification.
\end{abstract}


\end{frontmatter}

\section{Introduction}

According to the Global Cancer Statistics 2018 \citep{bray2018global}, stomach cancer was responsible for over one million new cases in 2018 with an estimated 783,000 deaths, making it the fifth most frequently diagnosed cancer and the third leading cause of cancer death in the world. Importantly, incidence rates are markedly elevated in Eastern Asia (e.g., Japan and Republic of Korea), whereas the rates in Northern America and Northern Europe are generally low and are equivalent to those seen across the African regions \citep{bray2018global}. However, a series of studies has shown that the incidence of signet ring cell carcinoma (SRCC) of stomach (a subtype of poorly cohesive carcinoma) has been slowly increasing, especially in the United States \citep{moran2014society, taghavi2012prognostic, henson2004differential}. The great majority of SRCC occurs in the stomach, with the rest arising in other organs (e.g., breast, gallbladder, pancreas, urinary bladder, and colon) \citep{yokota1998signet}.

SRCC is an invasive gastric adenocarcinoma and can be accompanied by diffuse growth of adenocarcinoma cells associated with a wide range of desmoplastic reactions, in particular when infiltrating into the submucosa or beyond \citep{nagtegaal20202019}. This type of growth is defined as diffuse cancer according to the Lauren classification \citep{LAURN1965}. In the early stage of the disease, intramucosal SRCC appears as layered cancer cells in the superficial portions of the mucosa without desmoplasia \citep{yamashina1986variant, bamba2001time, tatematsu1986gastric}. The typical signet-ring cells contain intracytoplasmic mucin that compresses the nucleus to the periphery of the cell wall, and glandular formations are rarely observed. Due to these morphological appearances, some of the SRCC cells often appear to mimic crushed oxyntic glands, crushed mucous neck cells, the goblet cells of the intestinal metaplasia, and gastric xanthoma (histiocytic aggregation) \citep{arnold2018atlas}. This makes SRCC more likely to be missed on routine histopathological diagnoses. False negatives have a detrimental impact on the quality and accuracy of the pathological diagnosis, and it should be addressed urgently. 

Computational pathology has been gaining momentum over the past decade, in particular due to the large increase in resources that allow the digitisation and processing of Haematoxylin and Eosin (H\&E) stained glass slides of surgical and biopsy specimens into Whole Slide Images (WSIs). Machine learning, in particular deep learning, has found many applications in computational pathology, such as cancer detection and classification, cell detection and segmentation, and gene mutation expression for a variety of organs and pathologies \citep{hou2016patch,madabhushi2016image, litjens2016deep, kraus2016classifying, korbar2017deep, luo2017comprehensive, coudray2018classification, wei2019pathologist, gertych2019convolutional, bejnordi2017diagnostic, saltz2018spatial, graham2019hover, campanella2019clinical, li2019weakly, wang2019rmdl, shi2020graph, syrykh2020accurate, kalra2020yottixel, iizuka2020deep, kanavati2020weakly}. 

Preparing a large fully-annotated training dataset for WSI cancer classification is a tedious and time-consuming task. This is because WSIs are extremely large, with heights and widths in the tens of thousands of pixels, as a result of being scanned at magnifications of x20 or x40 in order to reveal cellular-level details. 
The large image size makes it difficult to train and apply a CNN directly to WSIs due to GPU memory constraints. To bypass the computational constraints, the typically adopted approach is to divide the WSI into a set of fixed-sized tiles \citep{campanella2019clinical, coudray2018classification, liu2017detecting,  kanavati2020weakly}. Training of the CNN is done by using the resulting labelled tiles as input. Classification of a WSI is done by applying the CNN in a sliding window fashion, classifying the individual tiles, then aggregating all their classification outputs into a final WSI classification. The aggregation could be as simple as taking the maximum probability output of the tiles or using an RNN model \citep{campanella2019clinical, iizuka2020deep}. Obtaining a dataset of labelled tiles can either be done by asking pathologists to draw contours on WSIs or to classify pre-extracted, fixed-sized tiles. The latter requires pre-fixing the tile size and having pathologists classify millions of tiles. This is a tedious task. The former is preferable as the tile size can be modified later, and viewing the WSI provides context to pathologists and allows them to draw contours on large cancer infiltration areas; however, it can still be tedious especially with complex cancer infiltration patterns requiring annotations of individual cells. Once annotated, a single WSI can produce thousands of labelled tiles for training. A large dataset of labelled tiles is a requirement for fully-supervised learning.

On the other hand, weakly-supervised learning is an alternative approach and requires only weakly-labelled data \citep{zhou2018brief}. Given that diagnoses of WSIs are readily available from reports, additional annotations by pathologists are not required. Weakly-supervised learning methods, such as multiple instance learning (MIL) \citep{dietterich1997solving}, can operate directly on the WSIs by using the diagnoses as slide-level labels. This is a highly attractive solution.
One particular advantage of MIL is that it can reduce the labelling requirement. MIL was initially proposed in the context of drug discovery \citep{dietterich1997solving, maron1998framework}, and subsequently found many applications in computer vision \citep{babenko2010robust}, including histopathology classification and segmentation \citep{xu2014weakly,ilse2018attention,komura2018machine, cheplygina2019not, sudharshan2019multiple, li2019weakly, wang2019rmdl, campanella2019clinical, kanavati2020weakly}.
The caveat in histopathology applications, however,  is that the method tends to require a large training dataset of WSIs in order to work well. This has been demonstrated recently by \cite{campanella2019clinical} using a dataset of 44,732 WSIs to classify prostate cancer, basal cell carcinoma, and breast cancer metastases, with a reported AUC of about 0.98 on three test sets of about 1,500 WSIs each. They observed that at least 10,000 WSIs were necessary for training to obtain a good performance. Both weakly- and fully-supervised learning could be used on a dataset that has a combination of detailed cellular-level annotations and slide-level labels.

Only recently has SRCC detection been investigated \citep{li2019signet, lin2020decoupled}. \cite{li2019signet} set up the MICCAI DigestPath2019 challenge where one task was SRCC instance detection. A training dataset was made publicly available consisting of a total of 455 images (of which 77 had SRCC). The images were crops of size 2000x2000 pixels extracted from WSIs. A total of 12,381 instances of SRCC were manually annotated; however, the dataset still contains unannotated instances of SRCC. \cite{li2019signet} proposed a semi-supervised framework for SRCC detection where the goal was to train a deep learning network to detect individual SRCC instances using the combination of annotated and unannotated SRCC instances. The model was then evaluated on a test set consisting of 227 images (of which 12 had SRCC). The 1st runner up at the challenge proposed using a specialised loss \citep{lin2020decoupled} to separate the contribution of annotated and unannotated training samples resulting in an improvement in SRCC instance detection on the test set. Although there might be some interest from a research perspective in detecting all instances of SRCC in a specimen in order to calculate measurements, such as the karyoplasmic ratio or the degree of atypia, and study their correlations with outcomes. However, from a clinical perspective, all that matters is detecting whether a specimen has SRCC.

In this paper, our aim is the clinical application of detecting SRCC in WSIs. To this end, we developed and investigated deep learning models for the classification of SRCC in WSIs. We trained seven models using a combination of transfer learning, fully-supervised learning, and weakly-supervised learning. We used a training dataset consisting of a total of 1,765 WSIs of which 100 WSIs had an SRCC diagnosis. A group of pathologists non-exhaustively annotated individual cells suspected of SRCC in all of the 100 WSIs. We performed an investigation of different training methods in order to best understand which aspects contribute to obtaining a good SRCC WSI classification given the available data.

\section{Methods and Materials}

\begin{figure*}[!t]
    \centering
    \includegraphics[scale=.17]{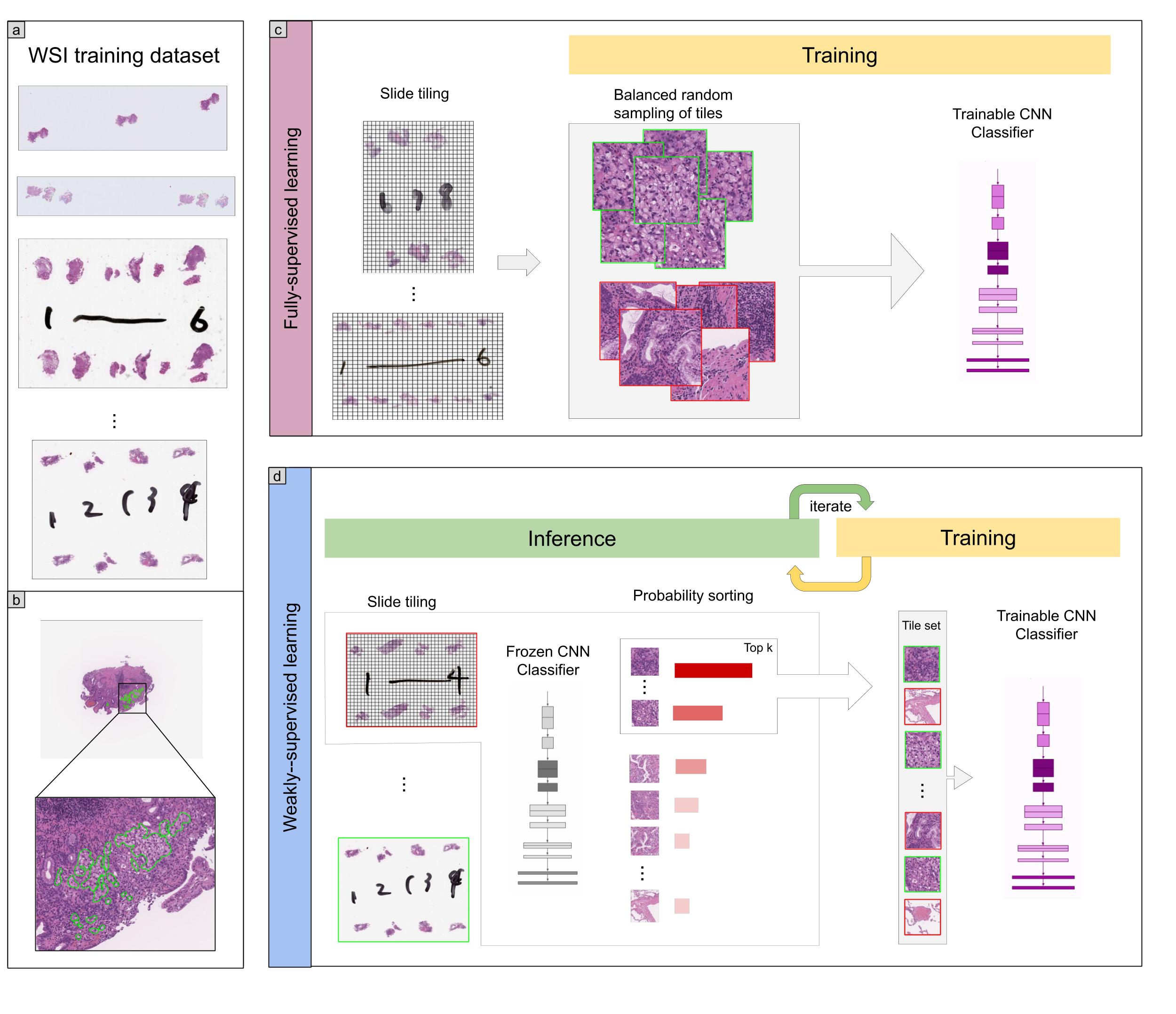}
    \caption{Overview of the training methods. (a) shows examples of biopsy WSIs in the training dataset. (b) shows an example of the SRCC annotations on WSIs. (c) shows an overview of the fully-supervised method where balanced batches of tiles are extracted from the WSI to train the CNN classifier. (d) shows an overview of the weakly-supervised method. The method alternates between two steps: inference and training. During inference a frozen CNN classifier is run in a sliding window fashion on each WSI and the top k tiles with the highest probabilities are placed into the training tile set. Once the training tile set reaches a certain size $T$, the training step is triggered.}
    \label{fig:training_methods_overview}
\end{figure*}

Our proposed method for SRCC WSI classification consists of using a CNN trained on tiles extracted from WSIs and using a combination of transfer learning, fully-supervised learning, and weakly supervised learning to train the models. Figure \ref{fig:training_methods_overview} provides an overview of the training methods.

\subsection{Problem formulation}

In histopathology, a pathologist diagnoses a WSI as having cancer if it is seen in any sub-region of the WSI; otherwise, it is diagnosed as not having cancer. This means that if a WSI with cancer were subdivided into a dense grid of smaller fixed-size tiles, then at least one of those tiles must have cancer, even though initially we do not know which tiles have cancer. If the WSI does not have cancer, then none of those tiles have cancer. This type of problem can be formulated generally with MIL. 
The MIL formulation adopts the concept of labelled bags that contain a collection of instances. A WSI $i$ is considered as a bag $H_i$ and any tile $j$ sampled from it is considered as a instance $x_{ij} \in H_i$. In the binary setting, a bag $H_i$ can either have a positive label ($y_i = 1$) or a negative label ($y_i = 0$). Similarly, an instance $j$ from bag $i$ can either have a positive label ($y_{ij} = 1$) or a negative label ($y_{ij} = 0$). The label of a bag $i$ is positive if at least one instance in the bag has a positive label.  This can formulated as $y_i = \smash{\displaystyle\max_j{(y_{ij})}}$.
As training is carried out on the instance level, the bag label $y_i$ is used to derive the labels $y^\prime_{ij}$ of the instances $x_{ij}$ for which the there are no labels.
The goal is then to train a model $f(x)$ that can classify all the instances. The MIL formulation allows training from data that has either purely labelled bags or a mix of labelled instances and bags, with one end of the extreme where only bags are labelled, and the other end of the extreme were all instances are labelled.
In this paper we are interested in classifying SRCC, so a positive label corresponds to a WSI having SRCC, and a negative label to the absence of SRCC.

\subsection{Training methods}

\subsubsection{Fully-supervised (FS) learning}
\label{sec:fs_method}
When we have labels for all the instances, there is no need to use the bag labels to derive the instance labels, and the MIL formulation becomes the classical fully-supervised (FS) learning method. The training dataset is $\{ (x_{ij}, y_{ij}) \}_{ i = 1,..., N, j = 1,...,N(i) }$, where $N$ is the number of WSIs and $N(i)$ is the number of tiles from the $i^{th}$ WSI. 
All positive WSIs would need to be annotated, potentially at cellular-level, such that labels are available for all the tiles. This task is tedious and time-consuming; however, it typically achieves the best performance, especially with a large labelled dataset. 

\subsubsection{Weakly-supervised (WS) learning} 
\label{sec:ws_method}
When we have labels only for bags or a mix of bags and instances, we can train the model using MIL. The training alternates between two steps: inference and training. Using the model trained so far, the inference step is used to extract a list of candidate tiles for training. During an epoch (one sweep through the entire dataset), we perform a balanced sampling (see Sec. \ref{sec:class_imbalance}) of tiles by randomly selecting in turn either a positive ($y_i=1$) or a negative ($y_i=0$) WSI. We then run inference on the WSI in a sliding window fashion and select the top $k$ tiles with the highest probabilities. This is done by sorting the probabilities in descending order and selecting the top $k$ instances.
From a positive WSI, the top $k$ correspond to tiles that the model is most confident that they contain cancer and their probabilities should be closer to 1. From a negative WSI, the top $k$ correspond to tiles that the model assigned the highest probabilities to, and they should be closer to 0. At each iteration, the top $k$ tiles are added to the set of training tiles. Once the size of the set reaches a certain threshold $T$, the set is shuffled and then fed into the model as batches for training. The model can alternate between inference and training many times within an epoch (if $T$ is less than the number of iterations in an epoch) or only once at the end of the epoch. At each iteration step, $y_i$ alternates between $0$ and $1$, and for selected instances that do no have a label, they are assigned the label $y^\prime_{ij} = y_i \forall x^\prime_{ij} \in \hat{H}^\prime_i(k)$, where

$$\hat{H}^\prime_i(k) = \operatorname*{argmax}_{H^\prime_i \subset H_i, |H^\prime_i| = k} \sum_{x_{ij} \in H_i'} f(x_{ij})$$

is the subset of top $k$ tiles.

\subsubsection{Weakly-supervised with fully-supervised pre-training} We can train the model by first training it with the FS method (Sec. \ref{sec:fs_method}), and then refining the model further by training it for additional epochs using the WS method (Sec. \ref{sec:ws_method}).

\subsubsection{Class imbalance}
\label{sec:class_imbalance} The training set was highly imbalanced, where WSIs with the negative class far outnumbered WSIs with the positive class (SRCC). To improve predictive performance on the positive class, we created a balanced sampler by over-sampling tiles from the positive class. This was done by having the tile sampler alternate from picking a fixed number of tiles from either a positive or a negative WSI. For FS, k tiles are picked randomly, whereas with WS, the top k tiles are picked based on their probabilities. The over-sampling ensures that all the negative WSIs are used for training during each epoch.

\subsection{Deep learning model}

We used the EfficientNet Convolutional Neural Network (CNN)  architecture \citep{tan2019efficientnet}, which has achieved state-of-the-art accuracy on computer vision datasets while having a  smaller number of parameters and a floating point operations per second (FLOPS) values that is an order of magnitude smaller compared to other existing architectures. 
The architecture uses compound scaling along width, depth, and image resolution of a baseline network, with mobile inverted bottleneck convolution (MBConv) as convolutional units. Different scales of EfficientNet have been trained on the ImageNet dataset \citep{deng2009imagenet}. We used the EfficientNet-B1 model architecture which has 7.8M parameters.

For transfer learning (TL), we initialised the weights of all the convolutional layers with the pre-trained weights on ImageNet. The final classification layer was a fully-connected layer with single output and a sigmoid activation function, and its weights were randomly initialised using the Glorot uniform initialiser \citep{glorot2010understanding}. During the first epoch, all the weights were frozen except for the weights of the final classification layer; this is so as to prevent random initial weight of the classification layer from destroying the pre-trained weights. After the first epoch, all the weights were unfrozen to become trainable.

\subsection{Tile extraction}
Tiles were extracted on the fly from the WSIs by direct indexing of locations without loading the entire WSI into memory. For a WSI, the locations were pre-computed as follows: first, we performed tissue detection by thresholding the image using Otsu's method; this step allowed eliminating a large portion of the white background and reducing unnecessary sampling of tile instances from the background. If annotation are available, then they could be used to further reduce the valid tissue sampling regions. Then, given a stride that allows subdividing the WSI into a grid, we extracted grid cell locations only from the valid tissue regions. These grid cells location were then used to extract tiles at the desired tile size and magnifications. For all the models, we used a fixed tile size of $224 \times 224$ pixels, and a stride during training of $112\times 112$ pixels.

As tiles were extracted from the WSIs, we applied data augmentation on the fly in the form of tile flips, $90^{\circ}$ rotations, translations, and colour shifts in order increase robustness and add regularisation effect to the training.
 
 \subsection{WSI classification}
 The models were trained as classifiers on the tile level; however, to obtain a WSI classification, the model was applied in a sliding window fashion using a stride of 112, then the WSI was assigned the label of the maximum probability of its tiles.

 \subsection{Heatmap visualisation}
 
We generated two types of heatmaps from the model using two methods: classification probability and Gradient-weighted Class Activation Mapping (Grad-CAM)  \citep{selvaraju2017grad}. The former consists in the tiling of the classification probability outputs by mapping each input tile's $1\times 1$ classification probability to a $\text{stride}\times \text{stride}$ pixels output tile. This can result in a blocky heatmap visualisation, especially with large strides. The smaller the stride, the more fine-grained the output; however, this comes at an increase in  prediction time. The latter, Grad-CAM, is a method that uses the gradients of the target output flowing into the final convolutional layer to produce a coarse localisation map. With the EfficientNet-B1 model, this produces a $7\times7$ output for a $224 \times 224$ input tile. Using a stride that is smaller than the input tile size, the outputs can be further smoothed by averaging the overlapping tiles; however, this too results in an increase in prediction time. Figures \ref{fig:tp_case} and \ref{fig:fp_cases} show examples of probability heatmaps with a stride of $112\times112$ pixels, whereas Fig. \ref{fig:Grad CAM} and \ref{fig:gradcam_digest} show Grad-CAM visualisations with a stride of $32\times 32$ pixels.

 \subsection{Implementation details}

The deep learning models were implemented and trained using TensorFlow \citep{tensorflow2015-whitepaper}. We used OpenSlide\citep{goode2013openslide} to read WSIs on the fly without pre-extracting all the tiles.
AUCs were calculated in python using the scikit-learn package \citep{scikit-learn} and plotted using matplotlib \citep{Hunter:2007}. The 95\% CIs of the AUCs were estimated using the bootstrap method \citep{efron1994introduction} with 1000 iterations. 
 
 \subsection{Datasets}

\label{sec:dataset}
\subsubsection{Hospital A and B}
For the present retrospective study, 2,824 cases of gastric epithelial lesions HE (hematoxylin \& eosin) stained specimens were collected from the surgical pathology files of Hiroshima University Hospital (Hospital A) and Tokyo IUHW Mita Hospital (Hospital B) after being reviewed by surgical pathologists. The experimental protocols were approved by the Institutional Review Board (IRB) of the Hiroshima University (No. E-1316) and International University of Health and Welfare (No. 19-Im-007). All research activities complied with all relevant ethical regulations and were performed in accordance with relevant guidelines and regulations of each hospital. Informed consent to use histopathological samples and pathological diagnostic reports for research purposes had previously been obtained from all patients prior to the surgical procedures at both hospitals and an opportunity for refusal to participate in research was guaranteed by an opt-out manner. 

The combined dataset obtained from both hospitals consisted of 2,824 WSIs of which were divided into sets of 1,765, 60, and 999 for training, validation, and test, respectively. The training set consisted of 100 SRCC, 571 other adenocarcinoma, and 1,094 non-neoplastic lesion, validation consisted of 20 SRCC, 20 other adenocarcinoma, and 20 non-neopastic lesions, and the test set consisted of 78 SRCC, 82 other adenocarcinoma and 839 non-neoplastic lesion. Both training and test sets were solely composed of endoscopic biopsy specimen WSIs. The 100 SRCC WSIs were manually annotated by a group of 11 surgical pathologists who perform routine histopathological diagnoses by drawing around the areas that corresponded to SRCC. The pathologists carried out detailed cellular-level annotations on cells that fit the description of SRCC cells as defined by the World Health Organization (WHO) classification of tumors (i.e., the following three tumor cell morphologies were adopted: (1) a cell with an intracytoplasmic cyst filled with acid mucin, giving the classical signet-ring appearance; (2) a tumor cell with eosinophilic cytoplasmic granules containing neutral mucin with a slightly eccentric nucleus; and (3) a tumor cell in which the cytoplasm is distended, with secretory granules of acid mucin appearing like a goblet cell) \citep{arai2019does, bosman2010classification}. The other adenocarcinomas training set included the following subtypes: tubular (tub), poorly differentiated (por) and papillary (pap) types which did not include SRCC cells in WSIs \citep{bosman2010classification}. The non-neoplastic training set included the following categories: ulcer, gastritis, regenerative mucosa, fundic gland polyp and almost normal gastric mucosa.
Each annotated WSI was observed by at least two pathologists, with the final checking and verification performed by a senior pathologist.

\subsubsection{DigestPath2019}
The DigestPath2019 data\footnote{https://digestpath2019.grand-challenge.org/} was obtained from the signet ring task of the DigestPath2019 grand challenge competition, part of the MICCAI 2019 Grand Pathology Challenge \citep{li2019signet}. We used the provided training dataset as a test set given that the classification labels were available. The dataset consisted of 455 images from 99 patients, of which 77 images from 20 patients contained SRCC. The provided images were 2000x2000 pixels crops extracted at a magnification of $\times40$ from WSIs. The original intended task of the challenge was to detect all instances of SRCCs; however, we only perform SRCC classification on the images. The size of the images was then adjusted based on the expected magnification of a given model.

\section{Experiments and results}

\subsection{Set-up}
We trained using three different training methodologies: fully-supervised (FS), weakly-supervised (WS), and fully-supervised pre-training followed by weakly-supervised (FS-WS). This resulted in seven different models: FS x5, FS x10, FS x20, FS w/o TL x10, WS x10, WS-noanno x10, and FS+WS x10.

For the FS method, we training the models using WSIs at three different magnifications $\times5$, $\times10$, and $\times20$. For the magnification at $\times10$, we trained using the FS method with and without transfer learning (w/o TL). During training the balanced tile sampler ensured that at least $k = 40$ tiles were randomly selected from each WSIs during a given epoch.

For the WS methods, we trained the models at a magnification of at $\times10$. In addition, we trained two versions of the model where in one version we only sampled from the annotated regions from the positive WSI, and in the other version we sampled tiles without using any of the annotations (WS-noanno). We used a top k value of $1$, and $T=128$, meaning that training was run multiple times during an epoch.

We evaluated the models on four test sets consisting of roughly 500 WSIs each: Hospital A (n=500, 23 SRCC, 28 other adenocarcinoma, 449 non-neoplastic lesions), Hospital B (n=499, 55 SRCC, 54 other adenocarcinoma, 390 non-neoplastic lesions), Hospital A \& B (n=508, 78 SRCC,  67 other adenocarcinoma and 363 non-neoplastic lesions), and DigestPath2019 (n=455, 77 SRCC). Hospital A \& B consisted of all the SRCC cases from both hospitals, with the remaining negative cases randomly chosen from either A or B to reach roughly 500 cases.

\subsection{Model hyperparameters}

All models were trained with the same hyperparameters.
We used the Adam optimisation algorithm \citep{kingma2014adam} with $beta_1=0.9$ and $beta_2=0.999$, and a learning rate of $0.001$ with a decay of $0.95$ every $2$ epochs. We used a batch size to 32. The performance of a given model was tracked on a validation set. We used an early stopping approach to avoid overfitting with a patience of 10 epochs, meaning that training would stop if no improvement was observed for 10 epochs past the lowest validation loss. The model with the lowest validation loss was chosen as the final model.

\subsection{Model evaluation}
\begin{figure}[!t]
\centering
\includegraphics[scale=.40]{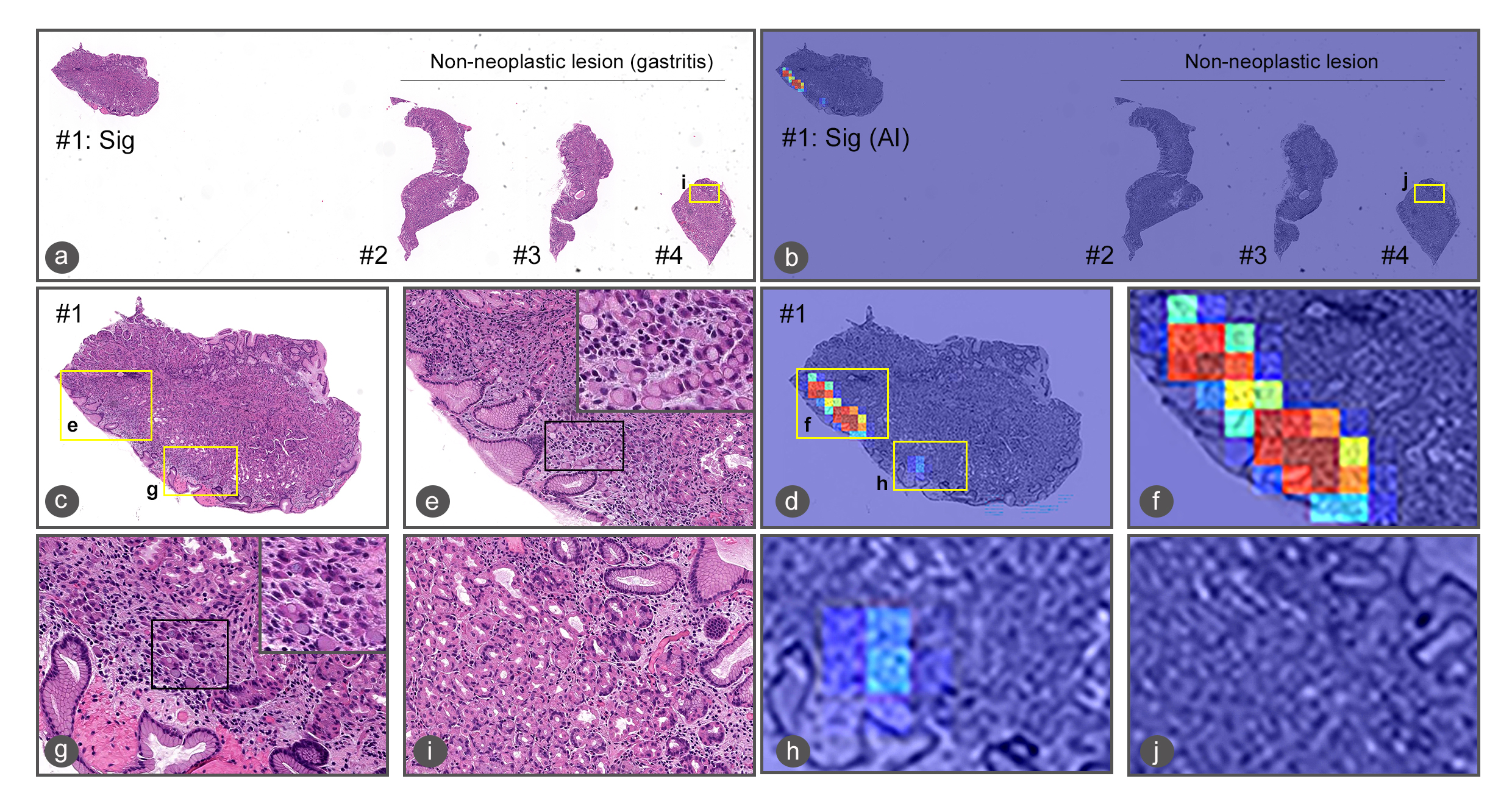}
\caption{Representative true positive case. There are four endoscopic biopsy fragments in this WSI (a). According to the pathological diagnostic report, \#1 is signet ring cell carcinoma and \#2-\#4 are gastritis (non-neoplastic lesion) (Figure 1a). When viewed under low magnification, highlighting is visible only in \#1 on heatmap image (b). When the highlighted area in \#1 is magnified (c), strong and low-signal areas are seen (d); a large number of signet ring cell carcinoma cells (e) are observed in the strong-signal area (f) and a small number of signet ring cell carcinoma cells (g) are seen in the low-signal area (Figure 1h). Enlargement of the tissue in \#4 confirms that it does not contain any signet ring cell carcinoma cells (Figure 1i \& 1j).}
\label{fig:tp_case}
\end{figure}

\begin{figure}[!t]
\centering
\includegraphics[scale=.45]{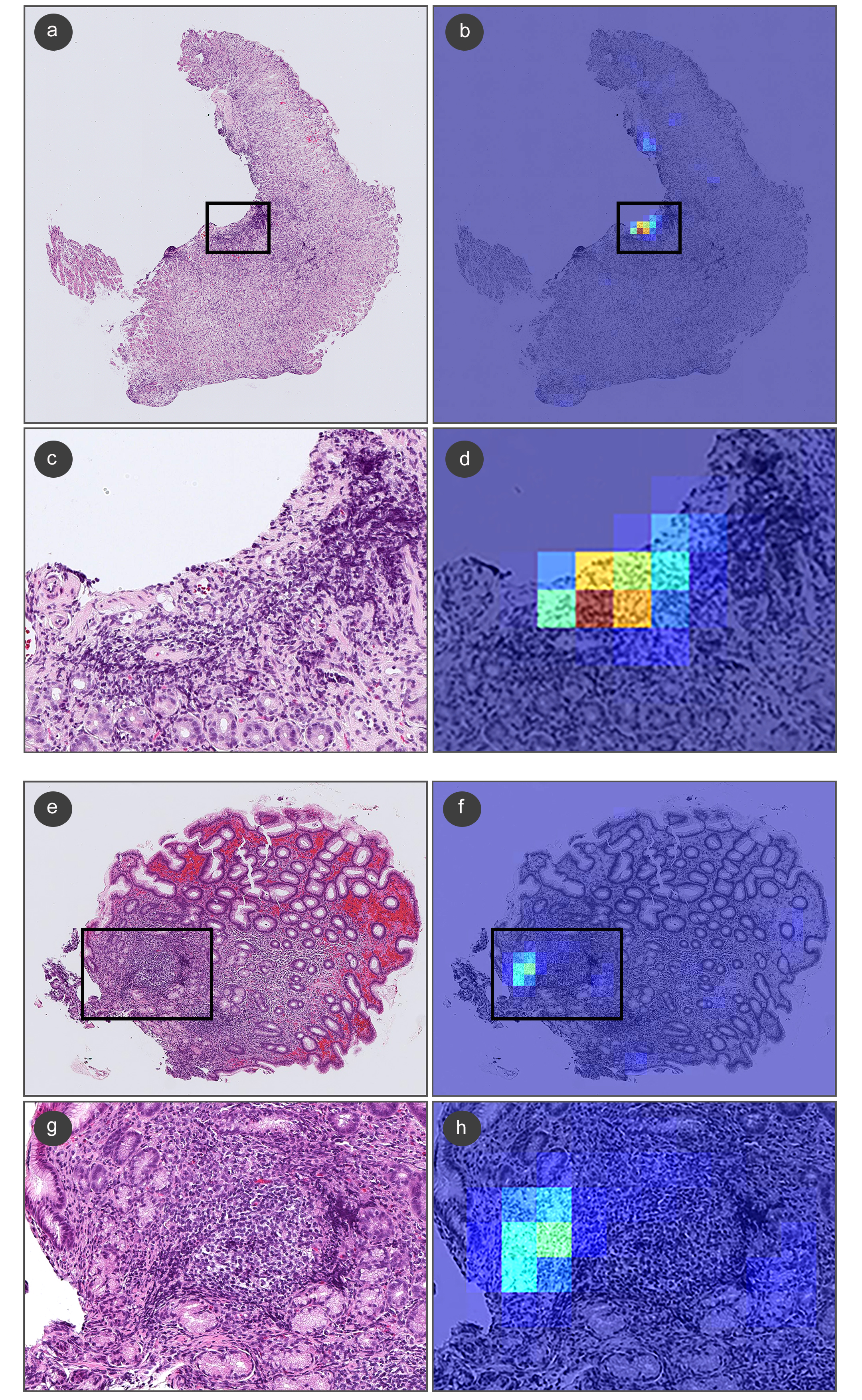}
\caption{Representative false positive cases. (a) is a case chronic gastritis (non-neoplastic lesion). (a-d) Pathologically, the false positives might be due to the lymphocytes being mixed around the smooth muscle cells and blood vessels of the muscularis mucosae and the nuclear density of the lymphocytes being similar to SRCC. 
(e) is a case of chronic gastritis (non-neoplastic lesion). (e-h) The false positive area includes pyloric glands disrupted by inflammation. Pathologically, the false positive area is suggested as a pyloric gland by comparison with other adjacent pyloric gland(s). However, on practical diagnosis, if such a finding is observed, additional investigation should be performed to confirm that it is a pyloric gland.}
\label{fig:fp_cases}
\end{figure}

\begin{figure*}[!t]
\centering
\includegraphics[scale=.2]{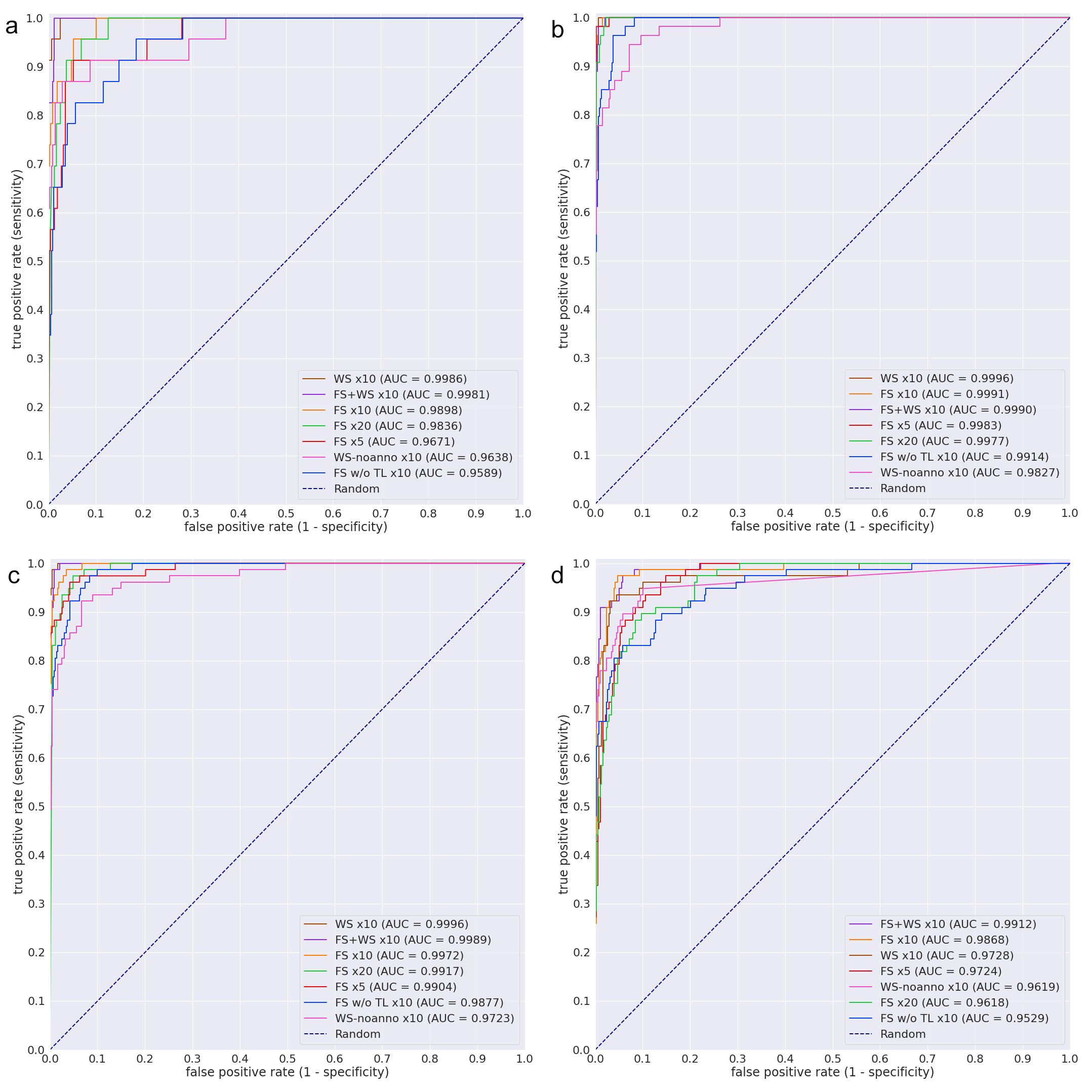}
\caption{ROC curves from the seven different models on the four test sets: (a) Hospital A, (b) Hospital B, (c) Hospital A \& B, and (d) DigestPath2019.}
\label{fig:roc_curves_testset}
\end{figure*}

We performed predictions on the WSIs of the test set by using a sliding window with an input tile size of $224\times 224$ pixels and a stride of $112\times112$ pixels. The WSI classification probability was obtained by max-pooling the probabilities of all of its tiles.
We computed the ROC curves and their corresponding AUCs as well as the log losses from all the models. Figure \ref{fig:roc_curves_testset} and Table \ref{tab:metrics} summarise the results on the test sets. Figure \ref{fig:tp_case} shows an example true positive classifications on four endoscopic biopsy fragments, whereas Fig. \ref{fig:fp_cases} shows examples of false positive classifications.

The models displayed good generalisation performance on the DigestPath2019 independent test set, which consisted of WSI crops obtained from a different source than the one used for training our models. We used the training set provided by DigestPath2019 as it was publicly available. We could not perform a direct comparison with the reported results of \cite{li2019signet} as the test set is not publicly available.

\begin{figure}[!t]
\centering
\includegraphics[scale=.45]{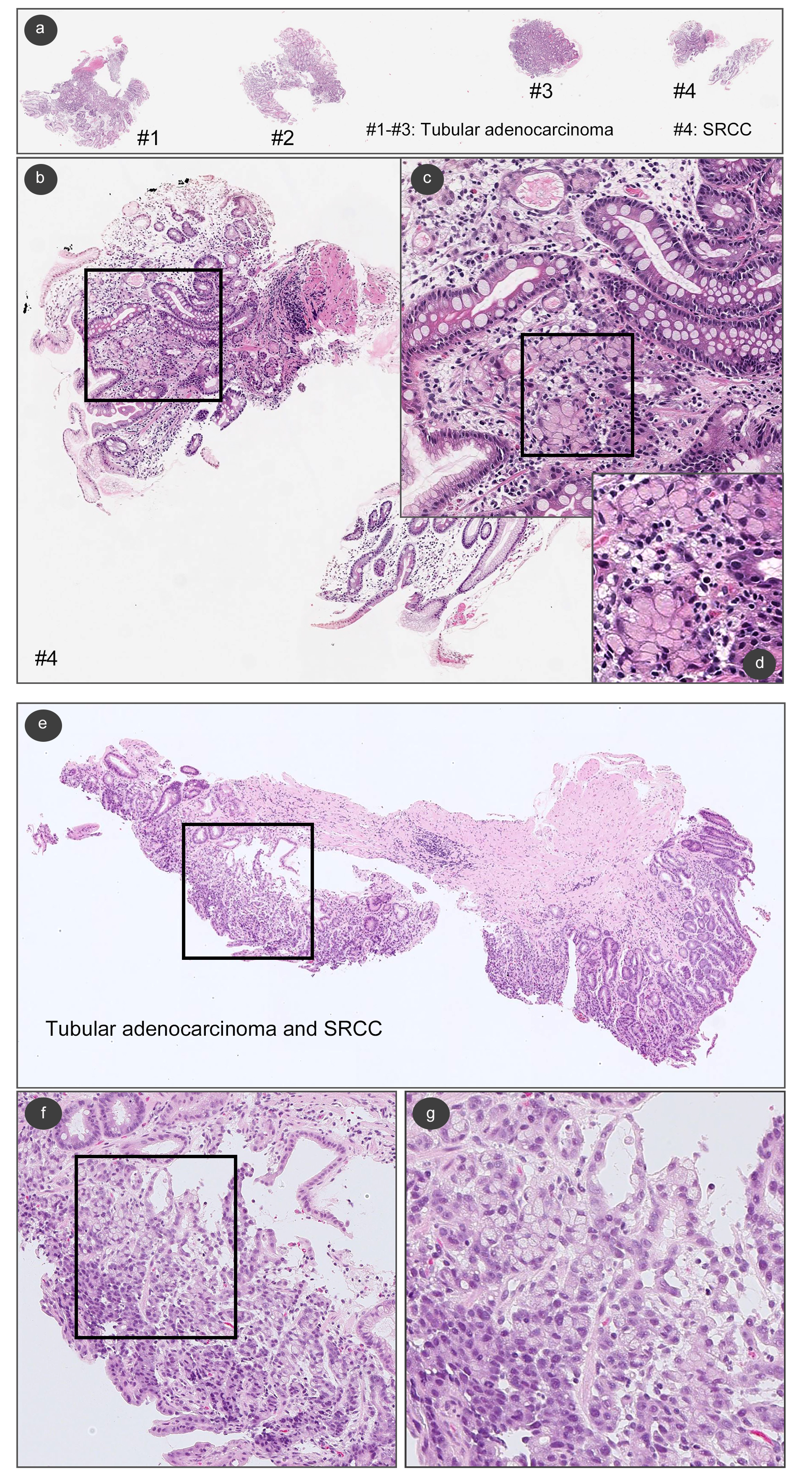}
\caption{Representative false negative cases. In (a) there are four endoscopic biopsy fragments (\#1-\#4). According to the pathological diagnostic report, (a) \#4 has SRCC. In the fragment of (b) \#4, a few SRCC cells were observed (c) at high magnification (d). (e) is endoscopic biopsy fragment. According to the pathological diagnostic report, this fragment has tubular adenocarcinoma and SRCC. When viewed under high magnification (f and g), SRCC cells were observed.}
\label{fig:fn_cases}
\end{figure}

\begin{figure}[!t]
\centering
\includegraphics[scale=.4]{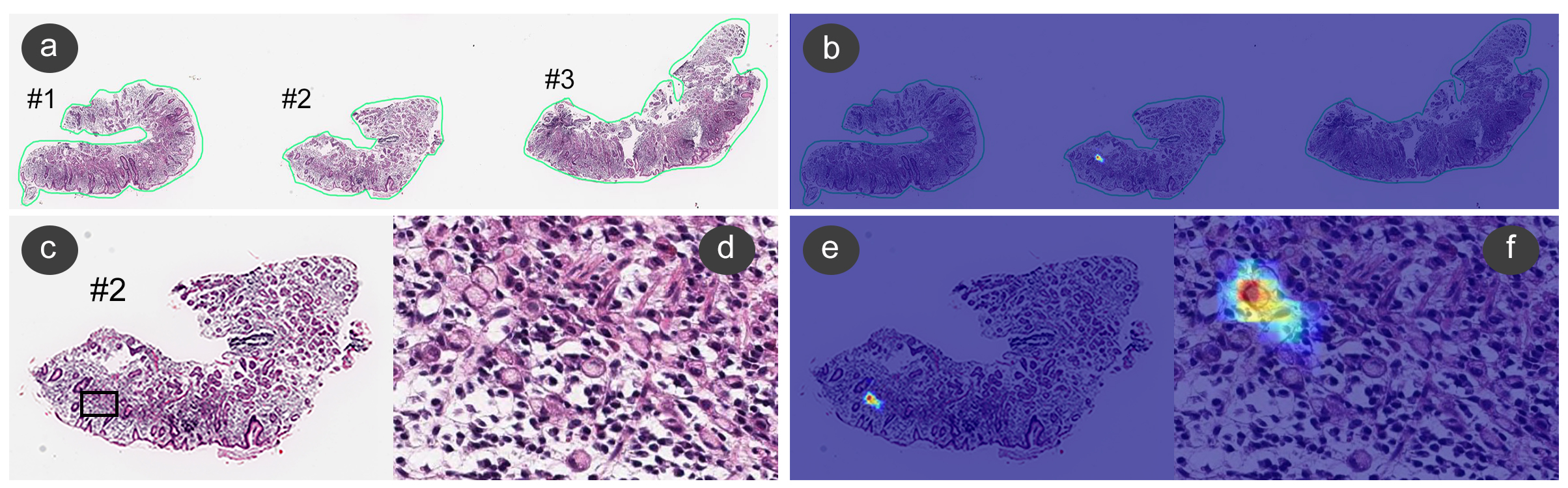}
\caption{Representative Grad-CAM heatmap image for true-positive detection of SRCC cells.  (a) shows non-neoplastic annotations (green lines) of gastric endoscopic biopsy specimens (\#1-\#3) by pathologists. Tissue fragments \#1 and \#3 are gastritis and \#2 has SRCC cells (a, c, d). Pathologists missed SRCC cells on 2 fragment (a). SRCC cells were visualized only in  fragment \#2 by Grad-CAM heatmap image (b). At high magnification, in fragment \#2, Grad-CAM hotspots (e, f) were overlapped with infiltrating area of SRCC cells (c, d). }
\label{fig:Grad CAM}
\end{figure}

\newcommand\setrow[1]{\gdef\rowmac{#1}#1\ignorespaces}
\newcommand\clearrow{\global\let\rowmac\relax}
\clearrow
\begin{table*}[!t]
\centering
\begin{tabular}{>{\rowmac}l>{\rowmac}l>{\rowmac}l>{\rowmac}l<{\clearrow}}
\toprule
                                         &               &                 ROC AUC &                log loss \\
\midrule
\multirow{6}{*}{Hospital A (n=500)} & FS x5 &  0.9671 (0.9315-0.9911) &  1.5891 (1.3425-1.9006) \\
                                         & FS x10 &  0.9898 (0.9781-0.9982) &   1.712 (1.5472-1.9224) \\
                                         & FS x20 &   0.9836 (0.969-0.9946) &  3.0657 (2.7967-3.3668) \\
                                         & \setrow{\bfseries}WS x10 &     0.9986 (0.9957-1.0) &    0.077 (0.0585-0.097) \\
                                         & FS+WS x10 &  0.9981 (0.9948-0.9998) &  0.3686 (0.3048-0.4469) \\
                                         & FS w/o TL x10 &  0.9589 (0.9265-0.9849) &  2.2022 (2.0085-2.3864) \\
                                         & WS-noanno x10 &  0.9638 (0.9213-0.9966) &  0.2862 (0.2139-0.3418) \\
                                         \\
\multirow{6}{*}{Hospital B (n=499)} & FS x5 &     0.9983 (0.9949-1.0) &  1.4644 (1.2338-1.7533) \\
                                         & FS x10 &     0.9991 (0.9974-1.0) &   1.1587 (1.0122-1.315) \\
                                         & FS x20 &   0.9977 (0.994-0.9997) &  3.0397 (2.8331-3.3458) \\
                                         & \setrow{\bfseries}WS x10 &     0.9996 (0.9986-1.0) &  0.0445 (0.0335-0.0589) \\
                                         & FS+WS x10 &      0.999 (0.9972-1.0) &  0.1786 (0.1338-0.2379) \\
                                         & FS w/o TL x10 &  0.9914 (0.9836-0.9965) &  1.7957 (1.6653-1.9641) \\
                                         & WS-noanno x10 &   0.9827 (0.9665-0.994) &  0.4758 (0.3922-0.5733) \\
                                         \\
\multirow{6}{*}{Hospital A \& B (n=508)} & FS x5 &  0.9904 (0.9781-0.9977) &   1.3494 (1.1385-1.579) \\
                                         & FS x10 &   0.9972 (0.994-0.9993) &  1.1106 (0.9592-1.2707) \\
                                         & FS x20 &  0.9917 (0.9844-0.9968) &   2.961 (2.7522-3.2976) \\
                                         & \setrow{\bfseries}WS x10 &     0.9996 (0.9988-1.0) &  0.0454 (0.0342-0.0596) \\
                                         & FS+WS x10 &  0.9989 (0.9971-0.9999) &  0.1861 (0.1372-0.2457) \\
                                         & FS w/o TL x10 &    0.9877 (0.978-0.994) &  1.6912 (1.5583-1.8581) \\
                                         & WS-noanno x10 &  0.9723 (0.9495-0.9887) &  0.4498 (0.3675-0.5476) \\
                                         \\
\multirow{6}{*}{DigestPath2019 (n=455)} & FS x5 &   0.9724 (0.956-0.9836) &  0.3667 (0.2477-0.5072) \\
                                         & FS x10 &  0.9868 (0.9739-0.9963) &  0.1584 (0.1037-0.2235) \\
                                         & FS x20 &  0.9618 (0.9396-0.9769) &  0.4678 (0.3869-0.5885) \\
                                         & WS x10 &  0.9728 (0.9486-0.9907) &   0.293 (0.2063-0.3798) \\
                                         & \setrow{\bfseries}FS+WS x10 &  0.9912 (0.9841-0.9974) &   0.0911 (0.0636-0.119) \\
                                         & FS w/o TL x10 &  0.9529 (0.9211-0.9741) &  0.2173 (0.1698-0.2808) \\
                                         & WS-noanno x10 &  0.9619 (0.9389-0.9877) &  0.4207 (0.2886-0.5324) \\
                                         \\

\bottomrule
\end{tabular}
\caption{ROC AUCs and log losses with their associated confidence intervals (CIs) for the four test sets: Hospital A, Hospital B, Hospital A \& B, and DigestPath2019.  }
\label{tab:metrics}
\end{table*}

\begin{figure*}[!t]
\centering
\includegraphics[scale=.12]{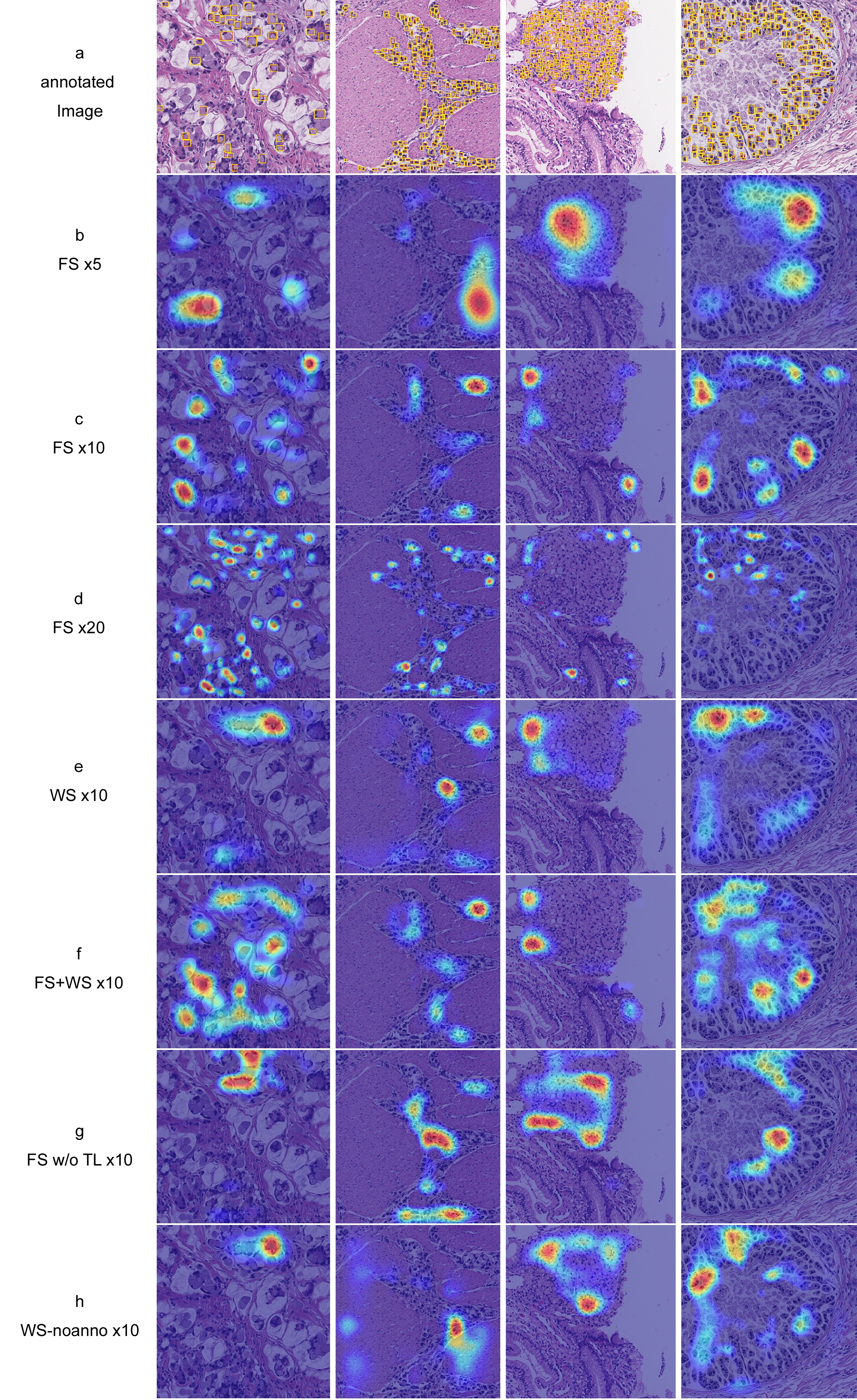}
\caption{Grad-CAM visualisation on positive images from the DigestPath2019 dataset. Row (a) shows four annotated images with yellow bounding boxes on SRCC instances. Rows (b-h) show the Grad-CAM outputs from the seven different models.}
\label{fig:gradcam_digest}
\end{figure*}

\section{Discussion}

In this paper we have presented a deep learning application for SRCC WSI classification. The models, based on the EfficientNet-B1 architecture, achieved high ROC AUC performance on four test sets, one of which originated from a different medical institution. We analysed the performance of different training methodologies and WSI magnifications. Results showed that a WS training method with WSIs at a magnification of x10 achieved the highest predictive performance.

The WS training method achieved a statistically significantly lower log loss compared to the FS method. This is most likely due to the WS method mainly training on tiles that have the highest probability from both the positive and negative WSIs, while the FS method trains on randomly sampled tiles. At each training iteration, the WS trains on the most confident tiles for the positive class and the most likely to be a false positive for the negative class. This prioritises the training on reducing the false positive rate, especially given that the WSI aggregation method is max-pooling. As a single false positive tile would result in a false positive classification for the WSI. Figure \ref{fig:gradcam_digest} shows Grad-CAM visualisation of the seven models on four positive images from the DigestPath2019. The models do no seem to pick up on the same areas. The FS+WS $\times10$ models picked up more SRCC cells than the WS $\times10$ model, which is most likely due to it having encountered more instances of SRCC tiles during the FS pre-training phase, with the WS training later serving to reduce the false positive rate.

For the WS method, guiding the sampling of the positive tiles from the annotated regions improved the predictive performance as compared to without using any of the annotations (WS x10 vs WS-noanno x10). This was to be expected given that there was only a small number of positive WSIs. Achieving a high predictive performance without any annotations that restrict the regions from which to sample requires a significantly larger dataset, as from the entire WSI of potentially thousands of tiles only one tile is selected for training. \cite{campanella2019clinical} observed that at least 10,000 are required to achieve a good performance.

Transfer learning was helpful in increasing predictive performance, given that the model trained without transfer learning (FS x10 w/o TL) mostly achieved the lowest performance on all four test sets. This was to be expected given that the training set contained a small number of positive WSIs (n=100).

The model trained at x20 has a higher false positive rate compared to the model trained at x10, and this is most likely due to the x10 model having more context information from the neighbouring tissues. In order to confirm an SRCC diagnosis, pathologists typically view a WSI at a low magnification (e.g., x4 or x5) and then at a higher magnification to check the cellular morphology.
It is more difficult for pathologists to distinguish between SRCC cells and mimicker cells (lymphocytes and histiocytes) if they are viewed in isolation without viewing the neighbouring tissues. The lack of context information from the neighbouring tissues could be the reason why the x20 model had a slightly lower predictive performance than the x10 model. 

The model trained at x5 similarly had a higher false positive rate compared to the model trained at x10; however, this is most likely due to the loss of cellular-level detail, making it harder to properly detect SRCC. Nonetheless, the model still achieved good performance. This result was particularly interesting to pathologists given that they would view the WSI with a magnification of at least x10 before confirming an SRCC diagnosis. 

An examination of some of the false positive cases showed that they were mostly due to cells exhibiting similar appearance to SRCC. In the chronic gastritis case in Fig. \ref{fig:fp_cases},  the nuclear density of the lymphocytes mimics the appearance of SRCC, which most likely led to the false positive. 
Figure \ref{fig:Grad CAM} shows a Grad-CAM visualisation of a case used as part of the validation set where a tissue fragment was incorrectly annotated as gastritis (non-neoplastic lesion). It was initially thought to be a false-positive case; however, another inspection by expert pathologists revealed that it is a true-positive detection of SRCC. It was missed by the pathologists performing the annotations potentially due to the presence of only a small number of SRCC cells within a background of chronic inflammatory cells infiltration that have some morphological similarities to SRCC cells, making them difficult to spot. Nonetheless, the models were able to make a correct detection.

As a certain element of randomness is involved when training the models, some of the variations in the predictive performance between the models could be attributed to it. However, the majority of the training methods achieved an acceptable high performance, signifying that it is possible to train an SRCC WSI classifier. One potential limitation is that we do not know the extent of how well the models generalise to WSIs from different source, given that most of the WSI test sets came from the same source as the training set. Nonetheless, the good performance on the DigestPath2019 dataset, even though it only consisted of WSI crops, is highly promising.
In addition, we do not know how well the models perform on challenging cases, such as intramucosal SRCC in-situ \citep{tsugeno2020histopathologic} and mimicker non-neoplastic cells like xanthoma\citep{drude1982gastric} cells, as neither the training or test sets contained any of these. 

\section{Conclusion}
In this study, we have developed a method of SRCC classification from WSI, and evaluated different training methods based that have different requirements on manual annotations. Annotating WSIs can be extremely tedious because of the massive size of WSI. We have shown that a weakly supervised method using minimal amounts of annotations can be used to train a WSI SRCC classification model with similar performance as a fully-supervised method.
Patients with SRCC tend to have poorer prognosis than patients with other types of gastric carcinoma \citep{liu2015clinicopathological, pernot2015signet}. However, recent studies have shown that the incidence of SRCC has been constantly increasing \citep{Bamboat2014, taghavi2012prognostic, henson2004differential}. Pathologists sometimes find SRCC more difficult to diagnose compared to other types of gastric carcinoma\citep{yamashina1986variant}. An AI model that can assist pathologists in detecting SRCC would be extremely beneficial as it can help them reduce diagnosis errors as well as potentially detect SRCC at an earlier stage and, as a result, significantly improve patient prognosis \citep{machlowska2019state}.

\section*{Author contributions statement}

F.K. and M.T. designed the studies; F.K., M.R., O.I. and M.T. performed experiments and analyzed the data; S.I. performed pathological diagnoses and helped with pathological discussion; K.A. provided pathological cases; F.K., S.I. and M.T. wrote the manuscript; M.T. supervised the project. All authors reviewed the manuscript.

\section*{Competing interests}

F.K., M.R., O.I. and M.T. are employees of Medmain Inc. S.I. and K.A. have no conflicts of interest. 

\section*{Acknowledgments}
We are grateful for the support provided by Professors Takayuki Shiomi \& Ichiro Mori at Department of Pathology, Faculty of Medicine, International University of Health and Welfare; Dr. Ryosuke Matsuoka at Diagnostic Pathology Center, International University of Health and Welfare, Mita Hospital; Dr. Naoko Aoki (pathologist). We thank the pathologists who have been engaged in the annotation work for this study.

\bibliographystyle{model2-names.bst}\biboptions{authoryear}
\bibliography{main}

\end{document}